\documentclass[polish]{webofc}
\usepackage{multirow}
\usepackage{tikz}
\usepackage{bbding}
\usepackage[varg]{txfonts}   % Web of Conferences font

\definecolor{kOrange}{RGB}{255,102,51}
\definecolor{kBlue}{RGB}{0,0,204}
\definecolor{kLightBlue}{RGB}{153,153,255}
\definecolor{kGreen}{RGB}{0,153,0}
\definecolor{kRed}{RGB}{204,0,0}
\definecolor{kCyan}{RGB}{51,204,204}
\definecolor{kMagenta}{RGB}{153,0,153}
\definecolor{kPink}{RGB}{204,0,102}
\definecolor{kBlack}{RGB}{0,0,0}

\newcommand{\tikzcircle}[2][kBlue]{\tikz[baseline=-0.5ex]\draw[#1,radius=#2] (0,0) circle ;}%
\newcommand{\W}{$\langle W\rangle$ }
\begin{document}
\title{News on mean pion multiplicity from NA61/SHINE}

\author{\firstname{Michał} \lastname{Naskręt}\inst{1}\thanks{\email{mnaskret@cern.ch}} for the NA61/SHINE Collaboration}

\institute{Institute of Theoretical Physics, University of Wroclaw}

\abstract{
NA61/SHINE is a large acceptance fixed target experiment at the CERN SPS which studies final hadronic states in interactions between various particles and nuclei~\cite{na61det}. The main topic of this contribution are preliminary results for mean negatively charged pion multiplicities $\langle \pi^{-} \rangle$ from central Ar+Sc and Be+Be collisions. The data were taken recently by the NA61/SHINE collaboration for a wide range of beam momenta. Measured rapidity distributions  $\frac{dn}{dy}$ were extrapolated to unmeasured regions to obtain total multiplicities $\langle \pi^{-} \rangle$. A new scheme to calculate the mean number of wounded nucleons $\langle W \rangle$ utilizing the EPOS MC model is described. Using data from other experiments, a comparison of $\frac{\langle \pi \rangle}{\langle W \rangle}$ for different collisions and beam momenta is discussed.
}

\maketitle

\section{$\pi^-$ rapidity distributions}
\label{rapidity}
The starting point of the analysis described herein are double differential spectra $\frac{d^2n}{dydp_{\text{T}}}$ of negatively charged hadrons, where $y$ and $p_{\text{T}}$ are rapidity and transverse momentum of partcles, respectively. Centrality was determined by selecting the 5\% of collisions with the smallest forward going energy as measured by the Projectile Spectator Detector (PSD)~\cite{na61det}.

The spectra were obtained from reconstructed tracks applying a series of quality cuts. In order to correct for trigger and reconstruction inefficiencies, one needs to apply a Monte Carlo correction. To this end, the EPOS MC~\cite{EPOS} is used in NA61/SHINE. A large statistics of ion collisions is generated and particles are accumulated in bins $n_{\text{gen}}^{i,j}$ in transverse momentum $p_{\text{T}}$ versus rapidity $y$. The generated data undergo the regular reconstruction procedure. Selecting negatively charged pions results in the distribution $n_{\text{sel}}^{i,j}$. The correction factor $c^{i,j}$ is then calculated as the ratio of the two Monte-Carlo generated spectra $c^{i,j}=n_{\text{gen}}^{i,j}/n_{\text{sel}}^{i,j}$. The final experimental spectra are obtained as $n^{i,j}=n^{i,j}_{\text{data}}c^{i,j}$.

In order to estimate the mean $\pi^-$ multiplicity in the full acceptance, one needs to extrapolate the experimental data to unmeasured regions. The extrapolation process consisted of two steps - extrapolation in transverse momentum $p_{\text{T}}$ for each bin of rapidity $y$ and extrapolation of ${dn}/{dy}$ in rapidity. For the latter a sum of two Gaussian functions was fitted, $g(y)=g_{\text{T}}(y)+g_{\text{P}}(y)$, where 

$$g_{\text{T}}(y)=\frac{A_0A_{rel}}{\sigma_0\sqrt{2\pi}}\exp\left(-\frac{(y-y_0)^2}{2\sigma_0^2}\right), g_{\text{P}}(y)=\frac{A_0}{\sigma_0\sqrt{2\pi}}\exp\left(-\frac{(y+y_0)^2}{2\sigma_0^2}\right)$$

In order to calculate the mean negatively charged pion multiplicity $\langle \pi^- \rangle$, the following formula was utilized:

$$\langle \pi^- \rangle = \int_{-4}^{y_{\text{min}}}g(y)dy + \sum_{y_{\text{min}}}^{y_{\text{max}}} dy\left(\frac{dn}{dy}\right)_{\text{extrapolated in} p_{\text{T}}}+\int_{y_{\text{max}}}^4 g(y)dy$$

Thus the final result is the sum over measured values of ${dn}/{dy}$ in the acceptance region and the integral over the Gaussian fits outside. Statistical uncertainties were calculated and systematic uncertainties were assumed to be 5\% based on the previous NA61 analysis of p+p collisions~\cite{antoni}.

\section{The mean number of wounded nucleons}
\label{wounded}
\begin{figure}[h]
  \centering
    \includegraphics[width=0.4\textwidth]{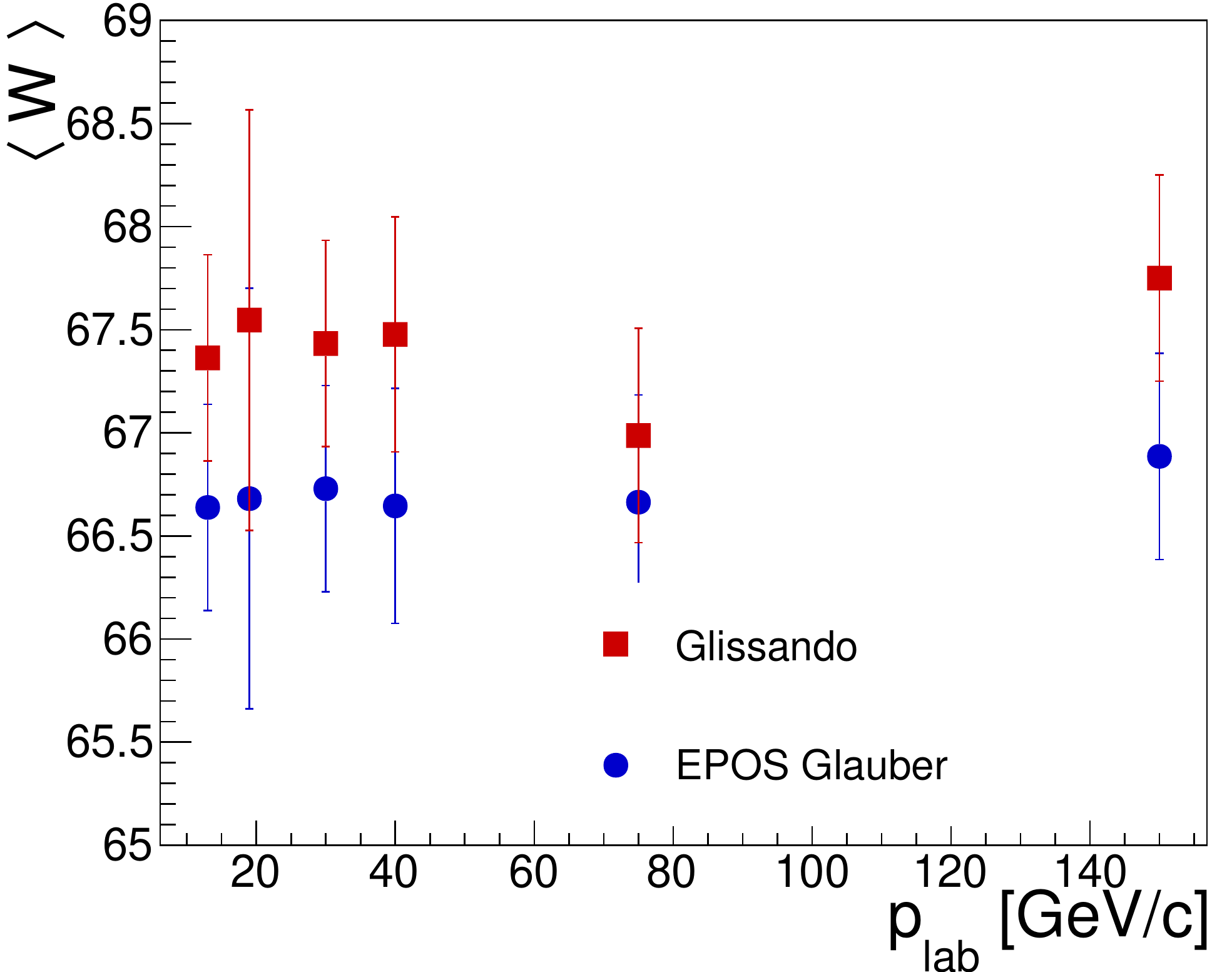}
  \caption{Average number of wounded nucleons $\langle W \rangle$ calculated by Glissando and EPOS "a la Glauber".}
  \label{fig:comparison}
\end{figure}

The number of wounded nucleons can not be measured experimentally in NA61/SHINE. It has to be calculated using Monte Carlo models. Two models were used to perform calculations - Glissando 2.73~\cite{glauber} based on the Glauber model and EPOS 1.99 (version CRMC 1.5.3)~\cite{EPOS} using a parton ladder model. Glissando provides values that are consistent with previous measurements and applicable to the wounded nucleon model~\cite{wounded}. EPOS, on the other hand, allows for more detailed centrality analysis and event selection. It is possible to reproduce Glauber-based \W values in EPOS and they are in good agreement with Glissando as shown in Fig.~\ref{fig:comparison}, where both - statistical and systematic uncertainties are calculated and combined. Therefore, Glauber-based EPOS values are used in later considerations.

\section{Results}
\label{results}

\begin{figure}
	\centering
	\includegraphics[width=0.4\textwidth]{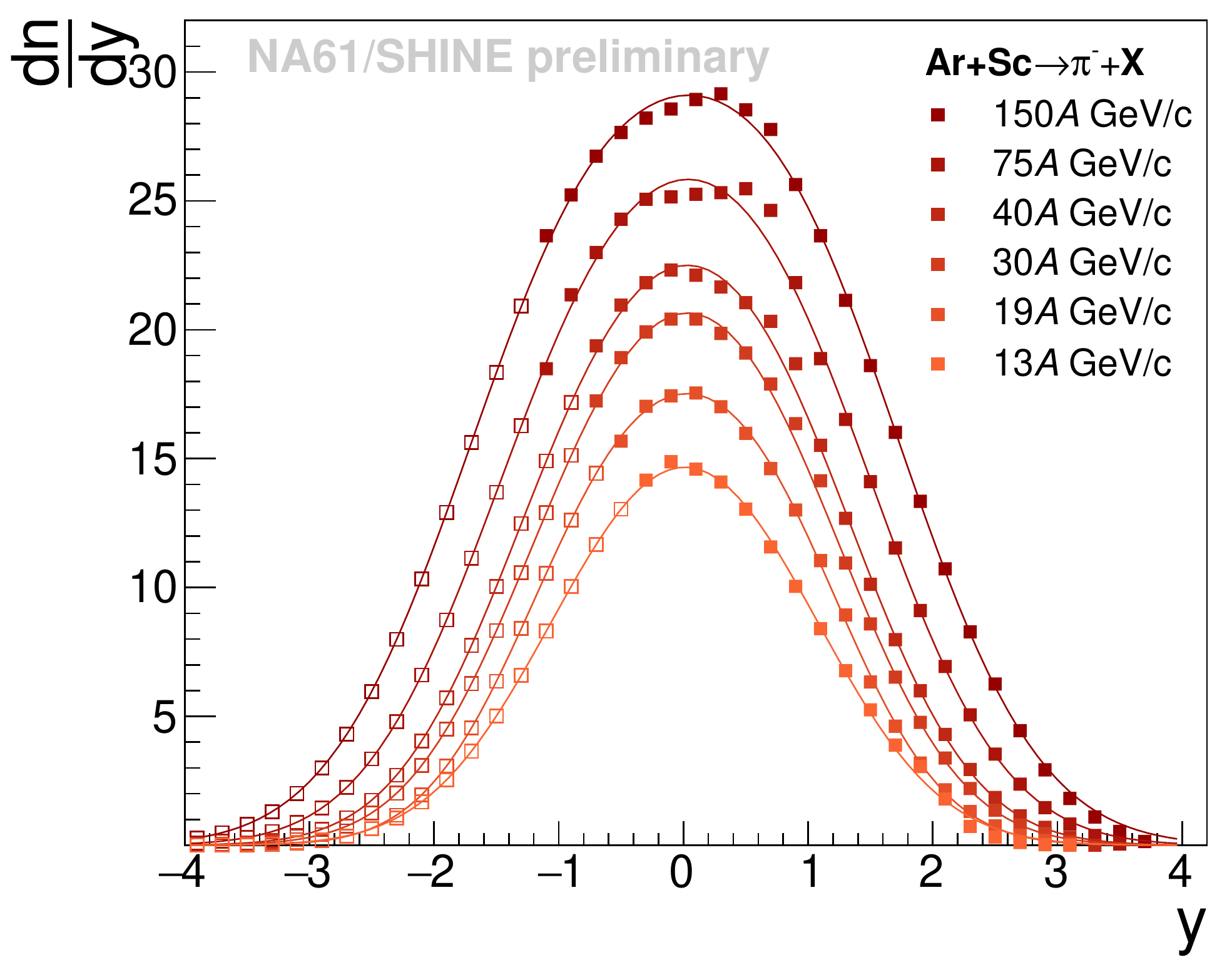}
	\includegraphics[width=0.4\textwidth]{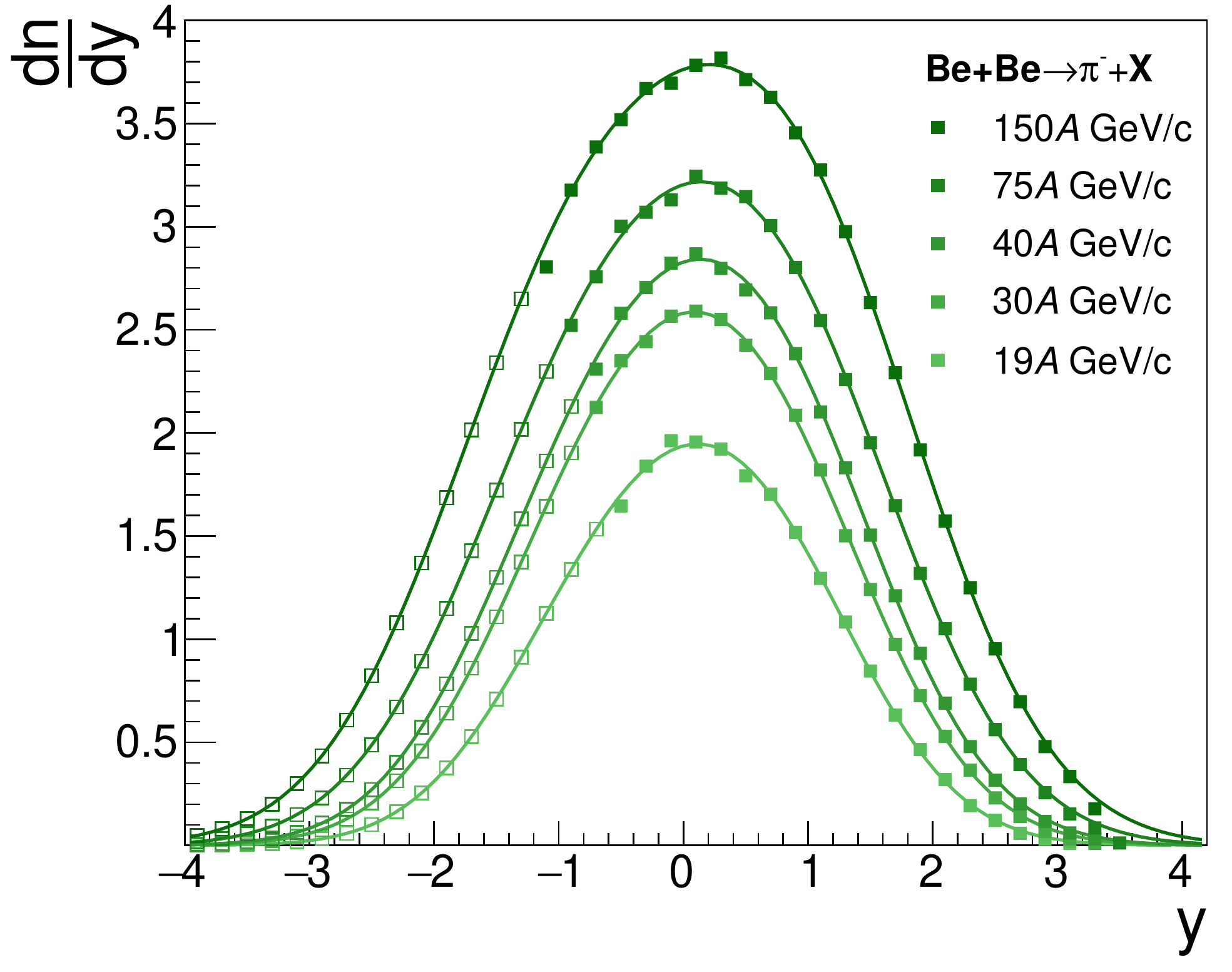}
	\caption{Preliminary $\pi^{-}$ rapidity spectra for the 5\% most central Ar+Sc and Be+Be collisions.}
	\label{fig:spectra}
\end{figure}

Preliminary $\pi^-$ rapidity spectra for Ar+Sc and Be+Be collisions are presented in Fig.~\ref{fig:spectra}

Preliminary results on $\langle \pi^-\rangle$ and $\langle W\rangle$ were calculated according to procedures described in sections~\ref{rapidity} and~\ref{wounded}. These are presented in Table~\ref{tab:ArSc} for Ar+Sc and Be+Be collisions.

\begin{table}
	\centering
	\begin{tabular}{l|cc|l|cc}
		\multicolumn{3}{c|}{Ar+Sc} & \multicolumn{3}{c}{Be+Be}\\
	  \hline
	  $p_{\text{lab}}$ [\textit{A} GeV/c] & $\langle\pi^-\rangle$ & $\langle W \rangle$ & $p_{\text{lab}}$ [\textit{A} GeV/c] & $\langle\pi^-\rangle$ & $\langle W \rangle$\\
	  \hline
	  13 & $38.46\pm 1.92$ & $66.63\pm 0.50$ & 20 & $5.32\pm 0.54$ & $10.99\pm 1.02$\\
	  19 & $48.03\pm 2.40$ & $66.68\pm 1.02$ & 30 & $7.61\pm 0.76$ & $10.86\pm 0.50$\\
	  30 & $59.72\pm 2.98$ & $66.72\pm 0.50$ & 40 & $8.75\pm 0.44$ & $10.86\pm 0.57$\\
	  40 & $66.28\pm 3.31$ & $66.64\pm 0.57$ & 75 & $10.98\pm 0.55$ & $10.83\pm 0.52$\\
	  75 & $86.12\pm 4.30$ & $66.66\pm 0.52$ & 158 & $14.32\pm 0.72$ & $10.79\pm 0.50$\\
	  150 & $108.92\pm 5.44$ & $66.88\pm 0.50$ & & &
	\end{tabular}
	\caption{Preliminary reslults on $\langle \pi^-\rangle$ and $\langle W\rangle$ for the 5\% most central Ar+Sc and Be+Be collisions.}
	\label{tab:ArSc}
\end{table}

In order to compare results obtained for different systems, an isospin correction should be taken into account. To this end phenomenological formulas are used
$$\langle\pi^-\rangle_{\text{N+N}}=\langle\pi^-\rangle_{\text{p+p}}+\frac{1}{3},\langle\pi^-\rangle_{\text{Au+Au}}^{\text{I}}=(\langle\pi^-\rangle_{\text{Au+Au}}+\langle\pi^+\rangle_{\text{Au+Au}})/2$$

\begin{figure}[h]
	\begin{minipage}[b]{0.85\textwidth}
  	\centering
 		\includegraphics[width=0.48\textwidth]{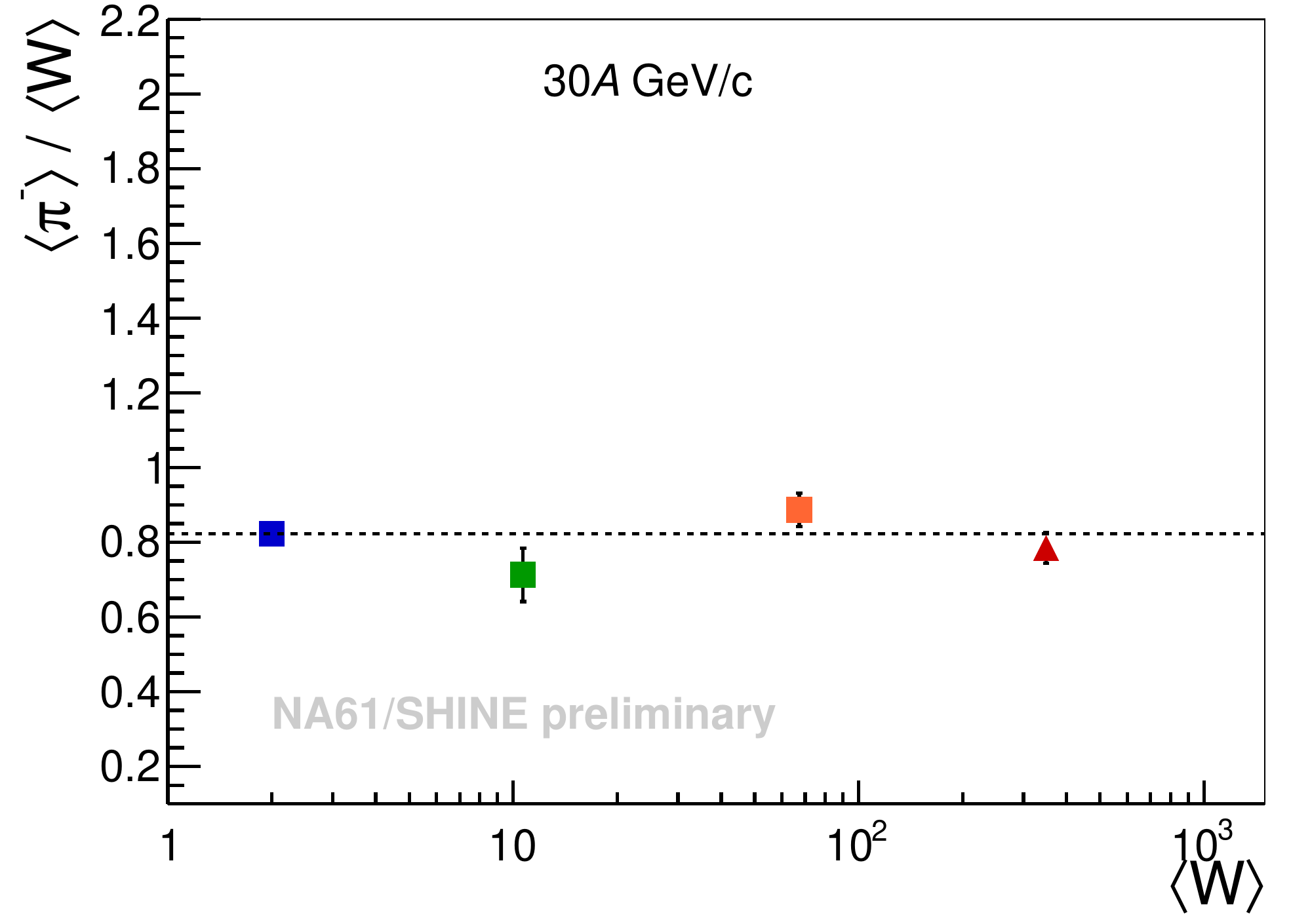}
  	\includegraphics[width=0.48\textwidth]{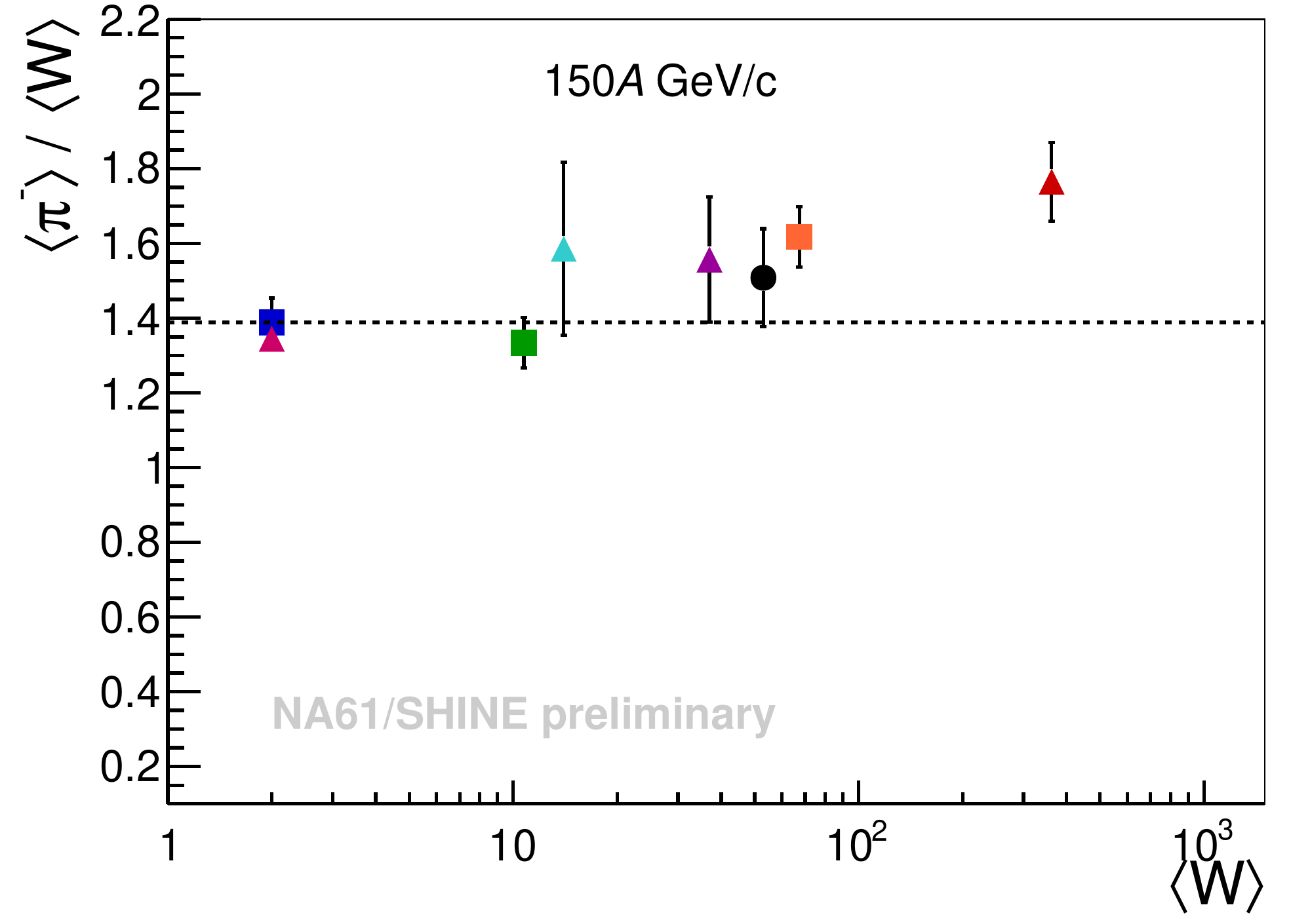}
	\end{minipage}
	\begin{minipage}[b]{0.14\textwidth}
\kern0pt
		\scriptsize
		NA61/SHINE
	  \begin{itemize}
	    \item[\textcolor{kOrange}{\SquareSolid}] Ar+Sc
	    \item[\textcolor{kGreen}{\SquareSolid}] Be+Be
	    \item[\textcolor{kBlue}{\SquareSolid}] N+N
		\end{itemize}
		NA49
	  \begin{itemize}
	    \item[\textcolor{kRed}{\TriangleUp}] Pb+Pb
	    \item[\textcolor{kMagenta}{\TriangleUp}] Si+Si
	    \item[\textcolor{kCyan}{\TriangleUp}] C+C
	    \item[\textcolor{kPink}{\TriangleUp}] N+N
		\end{itemize}
		NA35
	  \begin{itemize}
	    \item[\textcolor{kBlack}{\CircleSolid}] S+S
		\end{itemize}
	\end{minipage}
	\caption{Measurements of the $\langle\pi^-\rangle/\langle W \rangle$ ratio in nucleon-nucleon and central nucleus-nucleus collisions.}
	\label{fig:multWound}
\end{figure}

The correction is only applied to measurements where its effect is the strongest. This assumption is based on the compilation of the world data presented in~\cite{scaling} and the model presented therein. Where needed, the data is corrected for slight differences in beam momentum. Applying this correction one can plot the $\langle\pi^-\rangle/\langle W \rangle$ ratios for different systems for 30\textit{A} and 150\textit{A} GeV/c, see Fig.~\ref{fig:multWound}. Comparison of the preliminary Be+Be and Ar+Sc results with those from other systems suggest a monotonic increase with $\langle W \rangle$ at 150\textit{A} GeV/c and a constant value at 30\textit{A} GeV/c. 

The Fermi statistical model predicts a linear increase of $\langle \pi\rangle/\langle W\rangle$ with the Fermi energy measure $F=\left[(\sqrt{s_{\text{NN}}}-2m_{\text{N}})^3/\sqrt{s_{\text{NN}}}\right]^{1/4}$. An increase of the slope of $\langle\pi\rangle/\langle W\rangle$ versus $F$ -- the kink -- at the onset of deconfinement is predicted by the SMES~\cite{smes} model due to the larger number of effective degrees of freedom in comparison to the hadron resonance gas (HRG) model.

\begin{figure}
	\centering
	\begin{minipage}{0.5\textwidth}
    \centering
    \includegraphics[width=\textwidth]{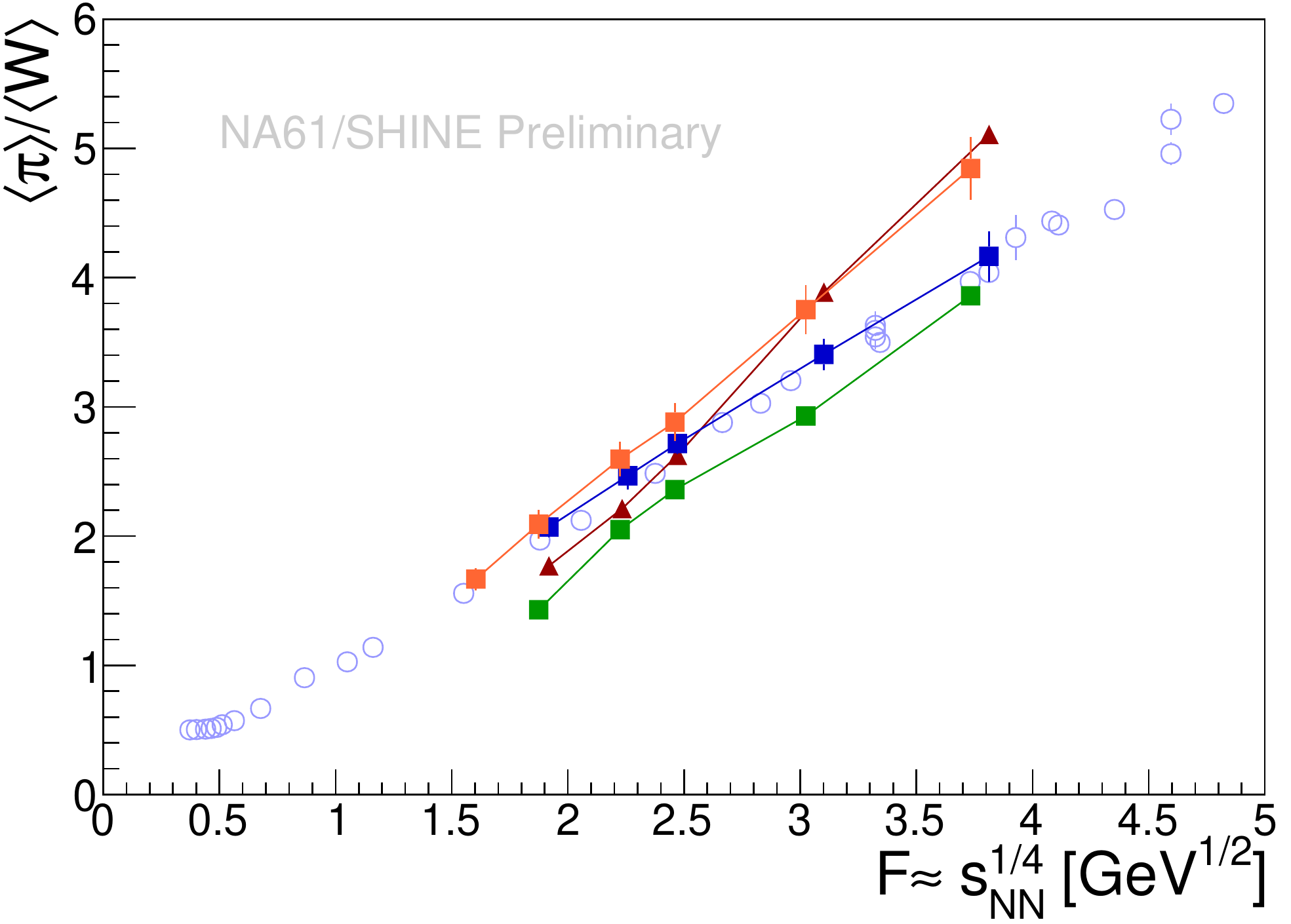}
	\end{minipage}
	\begin{minipage}{0.15\textwidth}
	\scriptsize
		NA61/SHINE
	  \begin{itemize}
	    \item[\textcolor{kOrange}{\SquareSolid}] Ar+Sc
	    \item[\textcolor{kGreen}{\SquareSolid}] Be+Be
	    \item[\textcolor{kBlue}{\SquareSolid}] N+N
		\end{itemize}
		NA49
	  \begin{itemize}
	    \item[\textcolor{kRed}{\TriangleUp}] Pb+Pb
		\end{itemize}
		WORLD
		\begin{itemize}
	    \item[\tikzcircle{3pt}] N+N
		\end{itemize}
	\end{minipage}
	\caption{The kink plot.}
	\label{fig:kink}
\end{figure}

As for the NA61 Ar+Sc, Be+Be and p+p data only the value of $\langle\pi^-\rangle$ was obtained, the multiplicities of $\langle\pi^+\rangle$ and $\langle\pi^0\rangle$ are approximated by multiplying the previously isospin asymmetry corrected $\pi^-$ multiplicities by a factor 3: $\langle\pi\rangle=3\langle\pi^-\rangle$. This approach is motivated by the fact that the NA61/SHINE acceptance is the largest for $\pi^{-}$. The energy dependence of the $\langle\pi\rangle/\langle W \rangle$ ratio is presented in Fig.~\ref{fig:kink}. The Be+Be results follow p+p whereas the Ar+Sc measurements follow Pb+Pb. No simple systematics are observed at low SPS energies. This might be caused by different physics processes and/or systematic bias in the estimate of $\langle W\rangle$. Full simulation of the fragmentation process and PSD response is needed in order to reduce the latter uncertainty.

\section{Conclusions}
Preliminary results on $\pi^-$ rapidity spectra and multiplicities in central Be+Be and Ar+Sc collisions at the CERN SPS are presented. Total multiplicities as function of collision energy and ratios with the average number of wounded nucleons were obtained. These new results are compared to previous measurements on nucleon-nucleon and nucleus-nucleus interactions. Results from Be+Be are similar to those from nucleon-nucleon collisions whereas results from Ar+Sc follow those obtained in Pb+Pb reactions at higher SPS energies.

This work was partially supported by the National Science Centre, Poland (grant 2015/18/M/ST2/00125).

\bibliography{bibl.bib}

\end{document}